\documentclass[12pt]{article}
\usepackage{amssymb}
\begin{document}
\begin{center}
QCD in the nuclear medium \\ 
and effects due to Cherenkov gluons

\bigskip

I.M. Dremin\footnote{email: dremin@lpi.ru}\\ 

{\it Lebedev Physical Institute, Moscow, Russia}

\end{center}

PACS numbers: 12.38

\begin{abstract}

The equations of in-medium gluodynamics are proposed. Their 
classical lowest order solution is explicitly shown for a color charge
moving with constant speed. For nuclear permittivity larger than 1 it
describes emission of Cherenkov gluons resembling results of classical
electrodynamics. The values of the real and imaginary parts of the nuclear 
permittivity are obtained from the fits to experimental data on the 
double-humped structure around the away-side jet obtained at RHIC.
The dispersion of the nuclear permittivity is predicted by comparing 
the RHIC, SPS and cosmic ray data. This is important for LHC experiments.
Cherenkov gluons may be responsible for the asymmetry of dilepton mass spectra 
near $\rho $-meson observed in the SPS experiment with excess in the low-mass 
wing of the resonance. This feature is predicted to be common for all resonances.
The "color rainbow" quantum effect might appear according to higher order 
terms of in-medium QCD if the nuclear permittivity depends on color.

\end{abstract}
 
\section{Introduction}

The collective effects observed in ultrarelativistic heavy-ion collisions at 
SPS and RHIC \cite{st, fw, PHENIX06, ph, ajit, ul} have supported the 
conjecture of quark-gluon plasma (QGP)
formed in these processes. The properties and evolution of this medium are
widely debated. At the simplest level it is assumed to consist of a set of
current quarks and gluons. It happens however that their interaction is quite
strong so that the notion of the strongly interacting quark-gluon plasma
(sQGP) has been introduced. Moreover, this substance reminds an ideal
liquid rather than a gas. Whether perturbative quantum chromodynamics (pQCD)
is applicable to the description of the excitation modes of this matter is
doubtful. Correspondingly, the popular theoretical approaches use either
classical solutions of in-vacuum QCD equations at the initial stage or 
hydrodynamics at the final stage of its evolution. However, it is surprising 
that no attempts to write down the equations of in-medium QCD, similar to
the very successful approach in electrodynamics, were done before the recent
paper \cite{qcdm} appeared.

The collective excitation modes of the medium may play a crucial role.
One of the ways to gain more knowledge about the excitation modes is to 
consider the propagation of relativistic partons through this matter.
Phenomenologically their impact would be described by the nuclear permittivity
of the matter corresponding to its response to passing partons. Namely this
approach is most successful for electrodynamical processes in matter.
Therefore it is reasonable to modify QCD equations by taking into account
collective properties of the quark-gluon medium. For the sake of simplicity
we consider here the gluodynamics only. The generalization to quarks is
straightforward.

The classical lowest order solution of these equations coincides with
Abelian electrodynamical results up to a trivial color factor. One of the most
spectacular of them is Cherenkov radiation and its properties. Now, Cherenkov
gluons take place of Cherenkov photons \cite{d1, ko}. Their emission in high 
energy hadronic collisions is described by the same formulae but with nuclear
permittivity in place of the usual one. 
Actually, one considers them as quasiparticles, i.e. quanta of the medium
excitations with properties determined by the permittivity. 

These formulae are used for fits of experimental data on the 
double-humped structure around the away-side jet obtained at RHIC. 
The values of the real and imaginary parts of the nuclear 
permittivity are determined. Comparing the RHIC, SPS and cosmic ray data one can
guess that the nuclear permittivity depends on energy. This leads to predictions
for future LHC experiments.

Beside the high energy region, the effects due to Cherenkov gluons might be 
noticed in asymmetry of the shapes of resonances passing through the nuclear
medium. This prediction is compared with experimental data of SPS. 

Another possible effect of "color rainbow" might appear if the nuclear 
permittivity differs for partons of different colors. The higher-order
non-linear terms of in-medium QCD equations are in charge of it.

\section{Equations of in-medium gluodynamics}

At the beginning let us remind the classical in-vacuum Yang-Mills equations
\begin{equation}
\label{f.1}
D_{\mu}F^{\mu \nu }=J^{\nu },
\end{equation}
\begin{equation}
\label{1}
F^{\mu \nu }=\partial ^{\mu }A^{\nu }-\partial ^{\nu }A^{\mu }-
ig[A^{\mu },A^{\nu }],
\end{equation}
where $A^{\mu}=A_a^{\mu}T_a; \; A_a (A_a^0\equiv \Phi_a, {\bf A}_a)$ are the 
gauge field (scalar and vector) potentials, the color matrices $T_a$ satisfy
the relation $[T_a, T_b]=if_{abc}T_c$, $\; D_{\mu }=\partial _{\mu }-ig[A_{\mu }, \cdot], \;\; 
J^{\nu }(\rho, {\bf j})$ is a classical source current, $\hbar=c=1$ and the 
metric tensor is $g^{\mu \nu }$=diag(+,--,--,--).

In the covariant gauge $\partial _{\mu }A^{\mu }=0$ they are written as 
\begin{equation}
\label{f.2}
%\framebox(1,1)  
\square A^{\mu }=J^{\mu }+ig[A_{\nu }, \partial ^{\nu }A^{\mu }+F^{\nu \mu }],
\end{equation}
where $\square $ is the d'Alembertian operator. It was shown \cite{kovc} 
(and is confirmed in what follows) that in this gauge the classical gluon 
field is given by the solution of the corresponding Abelian problem.

The chromoelectric and chromomagnetic fields are
\begin{equation}
\label{2}
E^{\mu}=F^{\mu 0 },
\end{equation}
\begin{equation}
\label{3}
B^{\mu}=-\frac {1}{2}\epsilon ^{\mu ij}F^{ij},
\end{equation}
or as functions of gauge potentials in vector notations
\begin{equation}
\label{4}
{\bf E}_a=-{\rm grad }\Phi  _a-\frac {\partial {\bf A}_a}{\partial t}+
gf_{abc}{\bf A}_b \Phi _c,
\end{equation}
\begin{equation}
\label{5}
{\bf B}_a={\rm curl }{\bf A}_a-\frac {1}{2}gf_{abc}[{\bf A}_b{\bf A}_c].
\end{equation}
The equations of motion (\ref{f.1}) in vector form are written as
\begin{equation}
\label{6}
{\rm div } {\bf E}_a -gf_{abc}{\bf A}_b {\bf E}_c = \rho _a,
\end{equation}
\begin{equation}
\label{7}
{\rm curl } {\bf B}_a-\frac {\partial {\bf E}_a}{\partial t} - gf_{abc}
(\Phi _b {\bf E}_c+[{\bf A}_b {\bf B}_c])= {\bf j}_a.
\end{equation}
                                                                  
The Abelian equations of in-vacuum electrodynamics are obtained from 
Eq. (\ref{f.2}) if the second term in its right-hand side is put equal to 
zero and color indices omitted. The medium is accounted if $\bf E$ is
replaced by ${\bf D} =\epsilon {\bf E}$ in $F^{\mu \nu} \;$, i.e. in Eq.
(\ref{2})\footnote{$\epsilon $ denotes the dielectric permittivity of the
medium. It is well known \cite{kel} that magnetic properties of
a substance are reproduced with the proper account of temporal and spatial
dispersion of $\epsilon $.}. Therefore the Eqs. (\ref{6}), (\ref{7}) in vector form are
most suitable for their generalization to in-medium case. 
The equations of in-medium electrodynamics differ from in-vacuum ones by
dielectric permittivity $\epsilon \neq 1$ entering there as
\begin{equation}
\label{f.3}
\bigtriangleup {\bf A}-\epsilon \frac{\partial ^2{\bf A}}{\partial t^2}=
-{\bf j},
\end{equation}
\begin{equation}
\label{f.4}
\epsilon (\bigtriangleup \Phi-\epsilon \frac{\partial ^2 \Phi}{\partial t^2})=
-\rho .
\end{equation}
The permittivity describes the matter response to the induced fields which is 
assumed to be linear and constant in Eqs. (\ref{f.3}), (\ref{f.4}). It is 
determined by the distribution of internal current sources in the medium.
Then external currents are only left in the right-hand sides of these equations. 

Now, the Lorentz gauge condition is
\begin{equation}
\label{f.5}
{\rm div } {\bf A}+\epsilon \frac {\partial \Phi }{\partial t}=0.
\end{equation}

The Lorentz invariance is lost if $\epsilon \neq 1$ in front of the second
terms in the left-hand sides. Then one has to deal within the coordinate system
where a substance is at rest. The values of $\epsilon $ are determined just
there. To cancel these requirements one must use Minkowski relations
between ${\bf {D, \; E, \; B, \; H}}$ valid for a moving medium \cite{llif}. 
It leads to more complicated formulae, and we do not use them in this paper.

The most important property of solutions of these equations is that
while the in-vacuum ($\epsilon = 1$) equations do not admit any radiation
processes, it happens for $\epsilon \neq 1$ that there are solutions of
these equations with non-zero Poynting vector.

Now we are ready to write down the equations of in-medium gluodynamics
generalizing Eq. (\ref{f.2}) in the same way as Eqs. (\ref{f.3}),
(\ref{f.4}) are derived in electrodynamics. We introduce the nuclear
permittivity and denote it also by $\epsilon $ since it will not lead
to any confusion. After that one should replace ${\bf E}_a$ in Eqs. (\ref{6}),
(\ref{7}) by $\epsilon {\bf E}_a$ and get:
\begin{equation}
\label{8}
\epsilon ({\rm div } {\bf E}_a-gf_{abc}{\bf A}_b {\bf E}_c)=\rho _a,
\end{equation}
\begin{equation}
\label{9}
{\rm curl } {\bf B}_a-\epsilon \frac {\partial {\bf E}_a}{\partial t} -
gf_{abc}(\epsilon \Phi _b{\bf E}_c + [{\bf A}_b{\bf B}_c])={\bf j}_a.
\end{equation}
The space-time dispersion of $\epsilon $ is neglected here.
 
In terms of potentials these equations are cast in the form:

\begin{eqnarray}
\bigtriangleup {\bf A}_a-\epsilon \frac{\partial ^2{\bf A}_a}{\partial t^2}=
-{\bf j}_a -
% \nonumber \\
gf_{abc}(\frac {1}{2} ( {\rm curl } [{\bf A}_b, {\bf A}_c]+
[{\bf A}_b {\rm curl } {\bf A}_c])+\frac {\partial }
{\partial t}({\bf A}_b\Phi _c)-  \nonumber \\
\epsilon \Phi _b\frac 
{\partial {\bf A}_c}{\partial t}- 
%\nonumber \\
\epsilon \Phi _b {\rm grad } \Phi _c-\frac {1}{2} gf_{cmn}
[{\bf A}_b[{\bf A}_m{\bf A}_n]]+g\epsilon f_{cmn}\Phi _b{\bf A}_m\Phi _n),  \label{f.6}
\end{eqnarray}

\begin{eqnarray}
\bigtriangleup \Phi _a-\epsilon \frac {\partial ^2 \Phi _a}
{\partial t^2}=-\frac {\rho _a}{\epsilon }+ 
%\nonumber  \\
gf_{abc}(2{\bf A}_b {\rm grad }\Phi _c+{\bf A}_b
\frac {\partial {\bf A}_c}{\partial t}-\epsilon 
\frac {\partial \Phi _b}{\partial t}
\Phi _c)+  \nonumber  \\
g^2 f_{amn} f_{nlb} {\bf A}_m {\bf A}_l \Phi _b.   \label{f.7}
\end{eqnarray}

If the terms with explicitly shown coupling constant $g$ are omitted, one gets
the set of Abelian equations which differ from electrodynamical equations
(\ref{f.3}), (\ref{f.4}) by the color index $a$ only. Their solutions are 
shown in the next section. The external current is ascribed to a parton fast 
moving relative to other partons "at rest". The crucial distinction between 
Eq. (\ref{f.2}) and Eqs. (\ref{f.6}),
(\ref{f.7}) is that there is no radiation (the field strength is zero in
the forward light-cone and no gluons are produced) in the lowest order solution 
of Eq. (\ref{f.2}) and it is admitted for Eqs. (\ref{f.6}), (\ref{f.7})
because $\epsilon $ takes into account the collective response (polarization)
of the nuclear matter. 

The omitted above terms are of the order of $g^3$ because the potentials and 
the classical current $J^{\mu }$ are linear in $g$. They can be 
taken into account as a perturbation. It was done in \cite{kov, ma} for
in-vacuum gluodynamics. For in-medium gluodynamics they are considered in 
Section 7 and Ref. \cite{adl}. 

\section{Cherenkov gluons as the classical lowest order solution of 
in-medium gluodynamics}

The classical solution of Eqs. (\ref{f.6}), (\ref{f.7}) immediately leads to
the notion of Cherenkov gluons at $\epsilon >1$ in analogy with Cherenkov
photons in electrodynamics. The unique feature is independence of the coherence 
of subsequent emissions by an external current on the time interval
between these processes. 

The problem of the coherence length for Cherenkov radiation was extensively 
studied \cite{tf, fr}. It was shown that the $\omega $-component of the field
of a current can be imitated by a set of oscillators with frequency $\omega $
situated along the trajectory. The waves from all oscillators add up in the
direction given by the Cherenkov angle $\theta $ independent on the length
of the interval filled in by these oscillators.
The phase disbalance $\Delta \phi $ between
emissions with frequency $\omega =k/\sqrt {\epsilon }$ separated by the
time interval $\Delta t $ (or the length $\Delta z=v\Delta t$) is given by
\begin{equation}
\label{f.9}
\Delta \phi =\omega \Delta t-k\Delta z\cos \theta =
k\Delta z(\frac {1}{v\sqrt {\epsilon }}-\cos \theta )
\end{equation}
up to terms which vanish for large distances between oscillating sources and
the detector. For Cherenkov effects the angle $\theta $ is
\begin{equation}
\label{f.10}
\cos \theta = \frac {1}{v\sqrt {\epsilon }}.
\end{equation}
The coherence condition $\Delta \phi =0$ is valid independent of $\Delta z $.
This is a crucial property specific for Cherenkov radiation only\footnote
{The requirement for $\Delta \phi $ to be a multiple of $2\pi $ (or a weaker
condition of being less or of the order of 1) in cases when Cherenkov condition 
is not satisfied imposes limits on the effective radiation length as it happens,
e.g., for Landau-Pomeranchuk or Ter-Mikaelyan effects.}. Thus the change of
color at emission vertices is not important if one considers a particular
$a$-th component of color fields produced at Cherenkov angle. Therefore the 
fields $(\Phi _a, {\bf A}_a)$ and the classical current for in-medium 
gluodynamics can be represented by the product of their electrodynamical 
expressions $(\Phi , {\bf A})$ and the color matrix $T_a$. As a result, one 
can neglect the "rotation" of color at emission vertices and use in the lowest 
order for Cherenkov gluons the well known formulae for Cherenkov photons just 
replacing $\alpha $ by $\alpha _SC_A$ for gluon currents in probabilities of 
their emission. Surely, there is radiation at angles different from the 
Cherenkov angle (\ref{f.10}). For such gluons one should take into account 
the coherence length and color rotation considering corresponding Wilson 
lines \cite{kwi}.

Let us remind the explicit Abelian solution for the current with velocity 
${\bf v}$ along $z$-axis
\begin{equation}
\label{f.11}
{\bf j}({\bf r},t)={\bf v}\rho ({\bf r},t)=4\pi g{\bf v}\delta({\bf r}-{\bf v}t).
\end{equation}

In the lowest order the solutions for scalar and vector potentials are
related so that
\begin{equation}
{\bf A}^{(1)}({\bf r},t)=\epsilon {\bf v} \Phi ^{(1)}({\bf r},t),   \hfill  \label{f.13}
\end{equation}
where the superscript (1) indicates the solutions of order $g$.

Therefore the explicit expressions for $\Phi $ suffice.
Using the Fourier transform, the lowest order solution of Eq. (\ref{f.4}) 
with account of (\ref{f.11}) can be cast in the form
\begin{equation}
\label{four}
\Phi ^{(1)}({\bf r},t)=\frac {g}{2\pi ^2\epsilon }\int d^3k
\frac {{\rm exp }[i{\bf k(r-v}t)]}{k^2-\epsilon ({\bf kv})^2}.
\end{equation}

The integration over the angle in cylindrical coordinates gives the Bessel
function $J_0(k_{\perp}r_{\perp})$. Integrating over the longitudinal 
component $k_z$ with account of the poles due to the denominator\footnote{These
poles are at work only for Cherenkov radiation!} and then over 
the transverse one $k_{\perp}$, one gets the following expression for the scalar 
potential \cite{ru}
\begin{equation}
\label{f.12}
\Phi ^{(1)}({\bf r},t)=\frac {2g}{\epsilon }\frac {\theta
(vt-z-r_{\perp }\sqrt {\epsilon v^2-1})}{\sqrt {(vt-z)^2-r_{\perp} ^2
(\epsilon v^2-1)}}.
\end{equation}

Here $r_{\perp }=\sqrt {x^2+y^2}$ is the cylindrical coordinate, $z$ is the
symmetry axis. The cone
\begin{equation}
\label{f.14}
z=vt-r_{\perp }\sqrt {\epsilon v^2-1}
\end{equation}
determines the position of the shock wave due to the $\theta $-function
in Eq. (\ref{f.12}). The field is localized within this cone. The Descartes
components of the Poynting vector are related according to Eqs. (\ref{f.12}),
(\ref{f.13}) by the formulae
\begin{equation}
\label{f.15}
S_x=-S_z\frac {(z-vt)x}{r_{\perp }^2}, \;\;\; 
S_y=-S_z\frac {(z-vt)y}{r_{\perp }^2},
\end{equation}
so that the direction of emitted gluons is perpendicular to the cone
(\ref{f.14}) and defined by the Cherenkov angle
\begin{equation}
\label{f.16}
\tan ^2\theta=\frac {S_x^2+S_y^2}{S_z^2}=\epsilon v^2-1,
\end{equation}
which coincides with (\ref{f.10}).

The higher order terms ($g^3$ ...) can be calculated using Eqs. (\ref{f.6}),
(\ref{f.7}) (see Section 7).

The expression for the intensity of the radiation is given by the Tamm-Frank
formula (up to Casimir operators)
\begin{equation}
\label{f.17}
\frac {dW}{dz}=4\pi \alpha_S\int \omega d\omega (1-\frac {1}{v^2\epsilon}).
\end{equation}
It is well known that it leads to infinity for constant $\epsilon $.
The $\omega $-dependence of $\epsilon $ (its dispersion) usually solves the
problem. 

For absorbing media $\epsilon $ acquires the imaginary part
$\epsilon=\epsilon _1+i\epsilon _2$.
The sharp front edge of the shock wave is smoothed.  The angular distribution 
of Cherenkov radiation widens. The $\delta $-function at the angle 
(\ref{f.10}), (\ref{f.16}) is replaced by the a'la Breit-Wigner angular shape 
\cite {gr} with maximum at the angle given by (\ref{f.10}), (\ref{f.16}) but 
with $\vert \epsilon \vert $ in place of 
$\epsilon $ and the width proportional to the imaginary part. 

The energy loss dW per the length dz is determined by the formula
\begin{equation}
\frac {dW}{dz}=-gE_z.    \label{eloss}
\end{equation}
In the lowest order
\begin{equation}
E_z^{(1)}=i\int \frac {d^4k}{(2\pi )^4}[\omega A_z^{(1)}({\bf k},\omega)-
k_z\Phi^{(1)}({\bf k},\omega)]e^{i({\bf k}{\bf v}-\omega)t},
\end{equation}
and in $k$-representation one gets in QCD
\begin{equation}
\Phi_a^{(1)}=2\pi gQ_a\frac {\delta (\omega -kv\zeta)v^2\zeta ^2}{\omega ^2
\epsilon(\epsilon v^2\zeta ^2-1)},
\end{equation}

\begin{equation}
A_{z,a}^{(1)}=\epsilon v\Phi_a^{(1)},
\end{equation}
\begin{equation}
\zeta=\cos \theta.
\end{equation}
The subscript of $\epsilon _a$ is omitted because there will be no summation 
on it even at higher order (see Section 7).

Thus, for ultrarelativistic radiating centers, the differential Cherenkov energy loss
 spectrum in the opaque medium per length $dz$ reads \cite{gr, GS03, dklv}
\begin{equation}\label{cherloss0}
 \frac{1}{\omega } \frac{d W}{dz \,d\omega \, d \phi \, d \cos \theta} =
 \frac{4 \alpha_S C_V}{\pi} \frac{\cos \theta (1-\cos^2\theta)
 \Gamma(\omega )}{\left(\cos^2\theta-\zeta (\omega )\right)^2+\Gamma^2(\omega)},
\end{equation}
where $\theta , \phi $ are polar and azimuthal angles with respect to the 
emitter propagation direction,
\begin{eqnarray}
 \zeta(\omega) &\; = &\; \frac{\epsilon_1(\omega)}{\epsilon_1^2(\omega)+
 \epsilon_2^2(\omega)}, \nonumber \\
 \Gamma(\omega) &\; = &\; \frac{\epsilon_2(\omega)}{\epsilon_1^2(\omega)+
 \epsilon_2^2(\omega)}, \nonumber \\
\end{eqnarray}
so that $\zeta(\omega)$ controls the location of the maximum and $\Gamma(\omega)$
controls the spreading around it. For $(\epsilon _2/\epsilon _1)^2\ll 1$, the
energy-dependent maximum of the differential spectrum (\ref{cherloss0}) is at
\begin{equation}\label{cosmax}
\cos^2 \theta_{\rm max} (\omega) \, \approx \, \frac{\epsilon_1(\omega)}
{\epsilon_1^2(\omega)+\epsilon_2^2(\omega)}.
\end{equation}

Without absorption, the potential (\ref{f.12}) is infinite on the cone.
With absorption, it is finite everywhere except the cone vertex and is inverse
proportional to the distance from the vertex. 
Absorption induces also longitudinal excitations (chromoplasmons) which are
proportional to the imaginary part of $\epsilon $ and usually small compared
to transverse excitations.

Several experimental observations may be explained as stemming from emission
of Cherenkov gluons.  Most accurate are the RHIC data with high statistics
about the two-humped structure of particle distributions around the away-side 
jets in central nucleus-nucleus collisions at ${\sqrt s}=200$ GeV per nucleon
\cite{st, fw, PHENIX06, ph, ajit, ul}. Their fit allowed to get values of the real and 
imaginary parts of the nuclear permittivity \cite{dklv}. The earlier cosmic
ray event \cite{apan}, which initiated the idea about Cherenkov gluons,
showed the smaller values of the nuclear permittivity \cite{drem1, IJMP}.
Quite small values of $\epsilon $ were also obtained from low-statistics 
samples of nuclei interactions at SPS energies \cite{vok, gho, aji}. This
indicates on energy dependence (dispersion) of $\epsilon $. The SPS data 
\cite{da} on asymmetry of the shapes of resonances passing through the nuclear 
medium can also be interpreted in terms of Cherenkov gluons \cite{dnec}.
Let us start with fits of the RHIC data. 

\section{Comparison with RHIC data}

In Eq.~(\ref{cherloss0}) the angle $\theta$ of emission of Cherenkov gluons
is measured with respect to the direction of propagation of the
radiating color current. In the experimental setup used at RHIC the special
trigger method has been used. Among all central Au-Au collisions at 200 GeV
per nucleon there were chosen only those where the partons of colliding
nuclei scattered at the angle about $\pi /2$. One of them passed a very
thin layer of a nucleus and produced the jet similar to those in pp-collisions.
Another one traversed through the whole nucleus and formed the away-side jet.
Namely around this jet the rings of hadrons were detected and interpreted as
originating from Cherenkov gluons. 
The partons initiating the hard away-side jets play the role
of the radiating color currents. Since the trigger is placed at $\pi/2$
to the colision axis of initial partons, the trigger parton is detected at
this angle. The away-side parton is also at this angle in the opposite 
direction if the energies of colliding partons are equal. It was shown
in \cite{IJMP} that the mismatch of these energies is unimportant because
the structure functions decrease fast for larger mismatch. The background
from gluons radiated by the aside moving partons is low and can result in
a slight widening of observed humps. Due to the forward-backward symmetry 
it does not influence the positions of the maxima. Thus, in the first 
approximation, for analytical estimates one can consider only those partons 
whose direction of propagation is orthogonal to the collision axis $z$.

In order to compare the spectrum (\ref{cherloss0}) with the
experimental data one should rewrite it in
terms of experimental laboratory polar and azimuthal angles $\theta_L$
and $\phi_L$. It is easy to see that for the above geometry
$$
    \cos \theta \, = \, \vert \sin \theta_L \cos \phi_L\vert ,
$$
for $\phi_L$ counted from the away-side jet.

The number $dN$ of emitted gluons per length $dz$ with energy and
angles within intervals $d\omega \, d\theta_L \, d\phi_L $ is
\begin{equation}\label{speclab}
 \frac{dN}{dz \, d\omega \, d \phi_L \, d \cos \theta_L} =\frac{4 \alpha_SC_V}{\pi}
 \frac{\vert \sin \theta_L \cos \phi_L \vert (1-\sin^2 \theta_L \cos^2 \phi_L)
 \Gamma(\omega)}{\left(\sin^2 \theta_L\cos^2\phi_L-\zeta(\omega)\right)^2+
 \Gamma^2(\omega)}.
\end{equation}

The azimuthal gluon distribution $dN/d\phi_L$ is obtained from 
Eq.~(\ref{speclab}) by integrating over $\theta_L$. This can
be done analytically:
\begin{equation}\label{cherloss}
 \frac{d N}{d\omega \, dl \, d \phi_L}= 4 \alpha_s C_V \frac{\Gamma}
 {\vert \cos\phi_L \vert} \left \{
 \frac{1}{\sqrt{\Gamma^2+\left( \cos^2 \phi_L - \zeta \right)^2} }
 \frac{1}{\sqrt{A^2+B^2}} \, \right. \times  
\end{equation}
$$
 \left[
 \left(  A + (1-\zeta)\frac{B}{\Gamma} \right) \sqrt{\frac{1}{2}
 \left(\sqrt{A^2+B^2}+A \right)} \right. -
$$
$$
 \left. \left. \left( B + (1-\zeta)\frac{A}{\Gamma} \right) \sqrt{ \frac{1}{2}
 \left(\sqrt{A^2+B^2}-A \right)} \right] - 1 \right \},  
$$
 where
\begin{eqnarray}
 A(\omega ,\phi_L) = \Gamma^2(\omega )+\zeta ^2(\omega )-\zeta (\omega )
 \cos^2 \phi_L , \;\;\;\;
 B(\omega ,\phi_L) = \Gamma(\omega ) \cos^2 \phi_L .
\end{eqnarray}

The gluon spectrum (\ref{cherloss}) reveals the double-humped structure. The
original flow of Cherenkov glue described by Eq.~(\ref{cherloss}) should, of 
course, transform into that of final hadrons. This can be done only with the 
help of Monte-Carlo models.

To describe properly the angular pattern observed in the flow of final 
hadrons in high energy nuclear collisions
\cite{st, ul, PHENIX06, ph} in terms of the Cherenkov gluon 
radiation one has to

{\bf (a)} Describe a kinematical pattern characterizing the initially produced 
hard partons serving as colored sources of gluon Cherenkov
radiation taking into account the experimental cuts on pseudorapidity and 
transverse momentum \cite{ul, ph}.

{\bf (b)} Write down a spectrum of Cherenkov gluon radiation in the opaque 
medium for the above-described color sources.

{\bf (c)} Describe a conversion of Cherenkov gluons into observed hadrons 
taking into account the experimental cuts on transverse momenta of final
hadrons \cite{ul, ph}.

These three steps are realized in a Monte-Carlo procedure. In total the model 
includes three parameters, described in more details below - two
related to the Cherenkov gluon radiation, the magnitudes of the real 
$\epsilon_1$ and imaginary $\epsilon_2$ parts of the permittivity of the
medium, and one related to rescattering of Cherenkov gluons in the process of 
hadronization $\Delta_\perp$. The values of these parameters are
fitted to achieve the best possible agreement with the experimental 
spectra \cite{ul, ph}.

The description of the initial hard color sources was performed through a 
Monte Carlo simulation of hard parton-parton scattering at RHIC energies
with PYTHIA \cite{PYTHIA}. This authomatically allowed to take into account 
possible mismatch of energies of initial colliding partons.
An initial pool of two-parton configurations 
resulting from these scatterings consisted of those in which the trigger
jet gave rise to a pion hitting the trigger intervals in rapidity, 
$|\,\eta|^{\rm \, tr}_{\rm \, star}\leq 0.7$ and $|\,\eta|^{\rm \, tr}_{\rm \,
phenix}\leq 0.35$, and transverse momentum, $3 \, {\rm GeV} \leq |\, 
{\bf p_\perp^{\rm lab}}| \leq 4 \, {\rm GeV}$ for STAR and 
$2 \, {\rm GeV} \leq |\, {\bf p_\perp^{\rm lab}}| \leq 3 \, {\rm GeV}$ for 
PHENIX \cite{ul, ph}. The fragmentation of gluon (quark) generating the 
trigger near-side jet into pions was described by standard fragmentation
functions \cite{A00}. 

The away-side jet gave rise to a pion hitting the rapidity intervals  
$|\,\eta|^{\rm \, aw}_{\rm \, star}\leq 1$ and $|\,\eta|^{\rm \, aw}_{\rm \,
phenix}\leq 0.35$, and transverse momentum, $1 \, {\rm GeV} \leq |\, 
{\bf p_\perp^{\rm lab}}| \leq 2 \, {\rm GeV}$ for STAR and 
$2 \, {\rm GeV} \leq |\, {\bf p_\perp^{\rm lab}}| \leq 3 \, {\rm GeV}$ for 
PHENIX \cite{ul, ph}. For simplicity we have considered only the dominating 
subset of events with gluonic away-side jet.

The angular distribution of emitted Cherenkov gluons is given by the expresion
(\ref{cherloss}).
Let us stress once again that in our Monte-Carlo procedure we consider a more 
general situation where the direction of the gluon generating the
away-side jet is fixed within each configuration satisfying the above-described 
trigger conditions imposed on the properties of the near-side jet in experiment.

To realize a  Monte-Carlo procedure of generating the Cherenkov spectrum we 
have to specify the functions $\epsilon_1(\omega)$ and
$\epsilon_2(\omega)$. Taking into account that the experimental cuts on 
(laboratory frame) intervals of transverse momenta of the away-side hadrons are very
strict, ${\bf \delta p_\perp^{\rm lab}} \vert \leq 1\,
{\rm GeV}$ both for STAR \cite{ul} and PHENIX \cite{ph} we can with
a good accuracy neglect the effects of dispersion and consider 
energy-independent $\epsilon_{1,2}={\rm const}$. The values of these constants are
determined through fitting the experimental data. Within this assumption the 
spectrum of produced Cherenkov glue is simply energy-independent, $d
N / d\omega \, dl={\rm const}$. Realistically the possibility of Cherenkov 
emission is restricted to some finite interval of energies $\omega \leq
\omega_{\rm max}$ so that
\begin{equation}
\frac{dN}{d \omega \, dl} \propto \theta \left(\omega_{\rm max}-\omega \right)\, ,
\end{equation}                                               
where $\omega_{\rm max}$ is the highest characteristic resonance excitation energy of the medium\footnote{For a discussion of interrelation between
Cherenkov gluons and hadronic resonances see \cite{dnec} and section 6.}. 
In our simulations 
we have chosen $\omega_{\rm max}=3.5\,{\rm GeV}$ in accordance with the upper
limits of experimental intervals. We have verified that
our results are in fact weakly sensitive to the exact value 
of $\omega_{\rm max}$.

The thus generated flow of Cherenkov glue should, of course, be transformed into that of final hadrons. There exist several phenomenological
schemes describing this conversion. In our case it is convenient to use a 
language of fragmentation functions $D^h_g(x,{\bf p_\perp}|\,Q^2)$ which
generically describe a probability for a gluon with energy $E$ and 
virtuality $Q^2$ to produce a hadron with the energy $xE$ and transverse
momentum ${\bf p_\perp}$ measured with respect to the direction of propagation of the initial gluon.

One is first tempted to ascribe the same shape to the spectrum of Cherenkov hadrons relying on the soft blanching hypothesis, equivalent to
assuming $D^h_g(x,{\bf p_\perp}|\,Q^2) \propto \delta(1-x)$ supported by the experimental evidence obtained in the physics of multiparticle
production from $e^+e^-\rightarrow $ jets. However, this does not lead to the fully satisfactory description of experimental data because of
predicting, in contradiction with the experimental data, that the probability 
of radiating Cherenkov glue at angles $\vert \, \phi _L\vert \geq
\pi/2$ with respect to the away-side jet is strictly zero. These "dead zones" appear because the Cherenkov gluon "halo" can not spread to angles
larger than $\pi /2$ to the direction of the emitter. It is clear that such "dead zones" will in fact be present for any fragmentation
mechanism that does not generate transverse momentum with respect to the direction of propagation of the initial Cherenkov gluon which in our case
corresponds to considering a ${\bf p_\perp}$-independent fragmentation 
function $D^h_g(x,|\,Q^2)$. We confine our consideration to fragmentation
to light hadrons because experimentally the share of other bosonic resonances 
is quite small (in particular, $\rho :\omega :\phi $=10:1:2
according to \cite{da}). In our analysis we used a simple parametrization of 
the fragmentation function $D^h_g(x,{\bf p_\perp}|\,Q^2)$:
\begin{equation}
 D^h_g(x,{\bf p_\perp}|\,Q^2) \propto (1-x)^3 \frac{1}{\sqrt{2 \pi \Delta_\perp^2}} \exp \left\{ -\frac{\bf p_\perp^2}{2 \Delta_\perp^2}
\right\}\, ,
\end{equation}
where the $x$-dependence is that of a gluon fragmentation function at the 
reference scale $Q_0=2\,{\rm GeV}$ \cite{A00}. Phenomenologically, the
parameter $\Delta_{\perp}$ accounts for transverse momenta acquired by pions
both due to the decay of intermediate resonances and because of rescattering 
processes.

Thus, we have a set of three parameters: $\epsilon_1, \epsilon_2 $ and a 
fraction $\Delta_{\perp} $.

The values of the parameters that provide the best description of experimental
data of STAR and PHENIX collaborations are shown in Table 1.

\bigskip

\begin{center}
{\bf Table 1}

\bigskip

\begin{tabular}{|c|c|c|c|c|}
  \hline
  Experiment & $\theta_{\rm max} $ & $\epsilon_1 $ & 
  $\epsilon_2 $ & $\Delta_{\perp} $ \\ \hline
  STAR & 1.04~rad & 5.4 & 0.7 & 0.7~{\rm GeV}\\ \hline
  PHENIX & 1.27~rad & 9.0 & 2.0 & 1.1~{\rm GeV}\\ \hline
\end{tabular}
\end{center}

\bigskip

Let us stress that a  difference in the fitted values of $\epsilon_{1}$ for STAR and PHENIX originates from different positions of angular maxima
$\theta_{\rm max}$ in the corresponding experimental data. (May be, it could indicate on necessity to take the dispersion into account.) However
these values are quite stable in both above approaches because they are defined by maxima positions but not by the humps widths.

The value of $\epsilon_{2}$, on the contrary, is influenced by the widths. 
Nevertheless, the ratio $(\epsilon_{2}/\epsilon_{1})^2\leq 0.05\ll 1$ stays
quite small in all cases.

The transverse smearing parametrized by $\Delta_\perp$ looks very reasonable
at the hadronic scale.

The resulting angular spectra for STAR and PHENIX are shown in Figs. 1 and 2 correspondingly. We see that the positions of the maxima (and
therefore the values of $\epsilon_{1}$) are quite stable to accounting for additional smearing on top of that in Eq.~(\ref{cherloss}). However, it
is indeed important in achieving a good description of the widths of humps in experimental data as seen in Figs. 1 and 2. The shape of humps in
the former "dead zone" determines the parameter $\Delta_\perp $, which, in its 
turn, influences $\epsilon _2$.

To conclude, the values of the real part of the nuclear permittivity 
$\epsilon_{1}$ found from the fit to experimental data of RHIC (albeit somewhat 
different in the two experimental sets) are determined with good precision. Their
common feature is that they are noticeably larger than 1. This shows that the density of scattering centers is quite large and the nuclear medium
reminds a fluid rather than a gas (for more details see \cite{IJMP}). The 
accuracy of the estimate of the values of $\epsilon_{2}$ is much less
but more important is the fact that they are rather small compared with 
$\epsilon_{1}$.

\section{Cosmic ray events and SPS data}

The RHIC experiments with high statistics used a special geometry of events 
and corresponding triggers to detect rings of hadrons around the away-side jets. 
Lower statistics data obtained at SPS and in cosmic rays also gave some 
indications on ring-like events. The very
first events with the ring-like structure were observed in non-trigger
cosmic ray experiments. Namely they initiated the idea about Cherenkov gluons
\cite{d1}. Two rings more densely populated by particles than their 
surroundings were noticed in the cosmic ray event \cite{apan} initiated by 
a primary with energy about 10$^{16}$ eV close to LHC energies. It is 
demonstrated in Fig. 3 where the number of produced particles is plotted as 
a function of the logarithm of the distance from the collision axis 
(proportional to pseudorapidity at small angles). It clearly shows two 
maxima. This event has been registered in the detector with nuclear and 
X-ray emulsions during the balloon flight at the altitude about 30 km. 
Approximately at the same time the similar event with two peaks was 
observed at $10^{13}$ eV \cite{arat}. Two colliding nuclei must give rise to 
two peaks. Some events with one peak (due to the limited acceptance of the 
installation?) were shown even earlier \cite{alex, masl}. The peaks were 
interpreted as effects due to Cherenkov gluons emitted by the forward and 
backward moving initial high energy partons. When the two-dimensional
distribution of particles was considered in the azimuthal plane (called 
as target diagram in cosmic rays experiments), this event revealed two
(forward and backward in c.m.s.) densely populated ring-like regions
within two narrow intervals of polar angles (corresponding to peaks in Fig. 3) 
but widely distributed in azimuthal angles for each of them. Therefore such 
events were called as ring-like events.

It was shown that the ring-like structure can be revealed by the event-by-event 
wavelet analysis \cite{dine, adk}.

The argument in favor of high energy Cherenkov gluons stems from the behavior
of the real parts of hadronic forward scattering amplitudes ${\rm Re}F(E, 0^o)$.
The intensity of the Cherenkov effect is proportional to the excess of the
permittivity $\epsilon $ or of the refractive index n over 1. They are related
by the formula $\epsilon = n^2$. The general relation of the scattering theory 
(see, e.g., \cite{go}) between the refractive index and the forward scattering 
amplitude $F(E, 0^o)$ states that
\begin{equation}                               
\Delta n={\rm Re}n-1 = \frac {2\pi N {\rm Re}F(E, 0^o)}{E^2}=
\frac {3m_{\pi }^3\nu _h }{2E^2}{\rm Re }F(E) =
\frac {3m_{\pi }^3\nu _h }{8\pi E } \sigma (E )\rho (E ).   \label{delta}
\end{equation}
Here $E$ denotes the energy, $N$ is the density of the scatterers 
(inhomogeneities) of the medium, $\nu _h$ is the number of scatterers within a 
single nucleon, $m_{\pi }$ the pion mass, $\sigma (E)$ the cross section 
and $\rho (E)$ 
the ratio of real to imaginary parts of the forward scattering amplitude $F(E)$. 
Thus the emission of Cherenkov gluons is possible only for processes with 
positive ${\rm Re} F(E)$ or $\rho (E)$. As follows from (\ref{delta}), the
values of $\Delta n$ can be completely different at low and high energies
because of different values of $N$ and ${\rm Re}F(E, 0^o)$ involved.
Unfortunately, we are unable to 
calculate directly in QCD ${\rm Re}F(E, 0^o)$ for gluons and have to rely
on analogies and our knowledge of properties of hadrons. The only experimental
facts we get about this medium are brought by particles registered at the final 
stage. They have some features in common which (one can hope!) are also
relevant for gluons as the carriers of the strong forces. 
These are the resonant behavior of hadronic amplitudes at rather low energies
and positiveness of ${\rm Re}F(E, 0^o)$ for all measured hadronic processes
at very high energies. Thus we wait for Cherenkov effects in the resonance 
region and at high energies.

The real parts of the resonance amplitudes are positive in their low-mass
wing. This fact will be extensively used in the next section. Here, it is
important that the energies of particles produced within the humps in RHIC 
data are rather low on the scale of initial energies. Even the energies 
of jets-parents are lower than 5 GeV. 
Therefore the whole effect at RHIC can be related to the low-energy region.

In cosmic ray events the energies of partons are high and the effect can be
ascribed to the high-energy behavior of ${\rm Re}F(E, 0^o)$.   
 
One of the most intriguing problems is to understand properly the fact 
that the RHIC and cosmic ray data
were fitted with very different values of the refractive index close to
3 and 1, correspondingly. This could be interpreted as due to the
difference in values of $x$ (the parton share of energy) and $Q^2$ (the
transverse momenta). It is well known that the region of large $x$ and $Q^2$
corresponds to the dilute partonic system. At low $x$ and $Q^2$ the density
of partons is much higher. 

The density of scattering centers at RHIC conditions is very high as estimated
from Eq. (\ref{delta}) \cite{IJMP}, about tens partons per proton volume.
Here one deals with rather low $x$ and $Q^2$. Therefore, one should 
expect the large density of partons in this region and high $\Delta n\sim 1$.
It is interesting to note that the two-hump structure disappears in RHIC data 
at higher $p_t$ where the parton density must get lower. It corresponds to 
smaller $n$ and $\theta $, i.e. humps merge in the main away-side peak.
  
In the cosmic ray event one observes effect due to leading partons with large 
$x$. Also, the experimentalists pointed out that the transverse momenta in this 
event are quite large. In this region one would expect for low
parton density and small $\Delta n\ll 1$. 

The ring-like events with values of $\epsilon $ close to but larger than 1 
were also found from the SPS non-trigger experiments \cite{vok, gho, aji}. 

Thus the same medium can be probably seen as a liquid or a gas depending 
on the parton energy and transferred momenta. This 
statement can be experimentally verified by using triggers positioned at 
different angles to the collision axis and considering different transverse 
momenta. In that way, the hadronic Cherenkov effect can be used as a tool to 
scan ($1/x, Q^2$)-plane and plot on it the parton densities corresponding
to its different regions. We discuss this problem also in Section 8.

\section{Asymmetry of shapes of resonances produced inside the nuclear medium} 

The low-energy Cherenkov gluons may be at the origin \cite{dnec} of another
interesting effect - the asymmetry of shapes of resonances created inside 
the nuclear medium. There exist numerous experimental data 
\cite{1, 2, da, 4, 5, Muto, 6, 7, 8} about
the in-medium modification of widths and positions of prominent vector-meson
resonances. They are mostly obtained from the shapes of dilepton mass and
transverse momentum spectra in nucleus-nucleus collisions. Such in-medium
effects were tied theoretically to chiral symmetry restoration a long
time ago \cite{9}.

A significant excess of low-mass dilepton pairs yield for $\rho $ meson over 
expectations from hadronic decays is observed in the high-statistics 
experiment \cite{da}. 

Several approaches have been advocated  for explanation of the excess.
Strong dependence of the parameters of the effective Lagrangian on the 
temperature and the chemical potential was assumed in \cite{10, 11}. The 
hydrodynamical evolution was incorporated in \cite{12} to describe the spectra.
The QCD sum rules and dispersion relations have been used \cite{13, 14} to
show that condensates decrease in the medium leads to both broadening and 
slight downward mass shift of resonances. The similar conclusions have been 
obtained from more traditional attempts using either the empirical scattering 
amplitudes with parton-hadron duality \cite{15, 15a} or the hadronic many-body 
theory \cite{16, 17, 18}.

In the latest approach, which pretends to provide the best description of 
experimental plots, the in-medium V-meson spectral functions are evaluated.
The excess of dilepton pairs below $\rho $-mass is ascribed to antibaryonic
effects. This conclusion is the alternative to more 
common ideas about the chiral restoration at high energies. It asks 
for some empirical constraints to fit the observed excess.

We proposed \cite{dnec} another possible source of low-mass lepton
pairs. Namely, the emission of Cherenkov gluons may provide a substantial 
contribution to the low mass region for any resonance. 

Qualitatively, the observed low mass excess of lepton pairs 
is easy to ascribe to the gluonic Cherenkov effect if one reminds that 
the index of refraction of any medium exceeds 1 within the lower wing of any 
resonance (the $\rho $-meson, in particular). 

This feature is well known in electrodynamics (see, e.g., Fig. 31-5 in 
\cite{fe}) where the atoms behaving as oscillators emit as Breit-Wigner
resonances when get excited. This results in the indices of refraction larger 
than 1 within their low-energy wings. In QCD, one can imagine that the nuclear 
index of refraction for gluons in the hadronic medium behaves similarly 
in the resonance regions. In classical electrodynamics, it is
the dipole excitation of atoms in the medium by light which results in the
Breit-Wigner shape of the amplitude $F(E, 0^o)$. In hadronic medium, there 
should be some modes (quarks, gluons or their preconfined bound states, 
condensates, blobs of hot matter...?) which can get excited by the impinging 
parton, radiate coherently if $n>1$ and hadronize at the final stage as 
hadronic resonances \cite{20, 21a}. 

The scenario, we have in mind, is as follows. Any parton, belonging
to a colliding nucleus, can emit a gluon. On its way through the nuclear medium
the parton excites some internal modes. Therefore it affects the 
medium as an "effective" wave which accounts also for the waves emitted by 
other scattering centers (see, e.g., \cite{go}). Beside incoherent scattering,
there are processes which can be described as the refraction of the initial 
wave along the path of the coherent wave. The Cherenkov effect is the induced 
{\it coherent} radiation by a set of scattering centers placed on the way of 
propagation of a gluon. That is why the forward scattering amplitude plays 
such a crucial role in formation of the index of refraction. At low energies 
its excess over 1 is related to the resonance peaks as dictated by the
Breit-Wigner shapes of the amplitudes (for the similar well known explanation 
in electrodynamics see, e.g., \cite{fe}). In experiment, usual resonances are 
formed during parton interactions and subsequent color neutralization process. 
Up to now there is no 
quantitative explanation of resonant amplitudes at the level of QCD-partons.
One can only admit some attractive forces acting within definite energy
intervals which lead to resonant behavior. In addition to these forces
between the individual partons the collective effects can appear in the nuclear
medium. As discussed above, there are several indications on such effects 
in RHIC experiments
like $J/\Psi $-suppression, azimuthal asymmetry $v_2$, jet quenching,
Cherenkov rings etc (see, e.g., \cite{st, fw, ph, ul}). Among them, Cherenkov 
gluons are distinguished as the result of the fully coherent reaction of the 
medium. Only in this case the phase of the amplitude does not depend on the 
time interval between emissions (see Section 4). The necessary condition $n>1$
is satisfied within the left-wing resonance region and coherent effect at low 
energies can be observed only there. Similarly to other partons, Cherenkov gluons
participate in formation of resonances contributing to their low-mass wing
(and, in particular, to dilepton spectra at these energies). Therefore these 
radiative effects can add to the ordinary ones (strings?) namely in this 
specific region. This contribution should be proportional to $\Delta n$ 
according to the theory \cite{tf}. The shock wave formed by Cherenkov gluons
\cite{qcdm} pushes out the hadronic states just with masses within the 
low-mass wings of resonances. Let us note that the common string 
description of the processes does not take into account collective effects. 
 
Thus, the ordinary Breit-Wigner shape of the cross section for resonance 
production must be modified by the coherent medium response. The easiest way
to observe this effect would be by measuring the dilepton mass spectra for
resonances. For all of them, they would acquire the additional term proportional 
to $\Delta n$ at masses below the resonance peak. 
According to Eq. (\ref{delta}) it must be 
proportional to the real part of the Breit-Wigner amplitude with 
the ratio of real to imaginary parts of Breit-Wigner amplitudes equal to
\begin{equation}
\frac {{\rm Re} F(M, 0^o)}{{\rm Im} F(M, 0^o)}=
\frac {m_{\rho }^2-M^2}{M\Gamma }\theta(m_{\rho }^2-M^2).
\label{reim}
\end{equation}
Here $M$ is the total c.m.s. energy of two colliding objects (the dilepton
mass), $m_{\rho }$=775 MeV is the in-vacuum $\rho $-meson mass.
Thus its $M$-dependence is well defined.
This term vanishes for $M>m_{\rho }$ because only positive $\Delta n$ lead
to the Cherenkov effect. Namely it describes the distribution of
masses of Cherenkov states. 

Therefore, the shape of the mass distribution near the $\rho $-meson can be 
described by the following 
formula\footnote{We consider only $\rho $-mesons here. To include other mesons,
one should evaluate the similar expressions.}
\begin{equation}
\frac{dN_{ll}}{dM}=\frac {A}{(m_{\rho }^2-M^2)^2+M^2\Gamma ^2}
\left(1+w\frac{m_{\rho }^2-M^2}{M^2 }\theta (m_{\rho}-M)\right)    \label{ll}
\end{equation}
The first term 
corresponds to the ordinary Breit-Wigner cross section with a constant
normalization factor $A$. According to the
optical theorem it is proportional to the imaginary part of the forward
scattering amplitude. The second term is due to the coherent response of the 
medium and its weight relative to ordinary processes\footnote{We assume it to 
be independent of $M$ because the relative probability of Cherenkov emission 
to ordinary processes can hardly change within this rather short interval 
of masses.} is described by the
only adjustable parameter $w$. As for any classical effect, one 
can not calculate the cross section for Cherenkov effect and determine the 
parameter $w$ theoretically. Therefore, we leave $w$ as
an adjustable parameter. The only hope is that it is not very small
because such an effect has been observed recently in high energy heavy-ion
collisions at RHIC (section 4) as the nimbus around away-side jets with 
rather high probability. 
Our fit below supports the assumption that the value of $w$ is quite noticeable
and corresponds to higher energy observations.

In these formulas, one should take into account 
the in-medium modification of the height of the peak and its width. In 
principle, one could consider $m_{\rho }$ as a free in-medium parameter as well.
We rely on experimental findings that its shift in the medium is small.
All this asks for some dynamics to be known. In our approach, it is
not determined. Therefore, first of all, we just fit the parameters $A$ 
and $\Gamma $ by describing the shape of the mass spectrum at $0.75<M<0.9$ GeV 
measured in \cite{da} and shown in Fig. 4. In this way we avoid any
strong influence of the $\phi $-meson. Let us note that 
$w$ is not used in this procedure. The values $A$=104 GeV$^3$ 
and $\Gamma =0.354$ GeV were obtained. The width of the in-medium 
peak is larger than the in-vacuum $\rho $-meson width equal to 150 MeV.
It agrees with previous findings of widening of in-medium $\rho $-mesons. 

Thus the low mass spectrum at $M<m_{\rho }$ depends only on a single parameter
$w$ which is determined by the (theoretically unknown) relative role of 
Cherenkov effects and ordinary mechanism of resonance production. It is clearly seen
from Eq. (\ref{ll}) that the role  of the second term in the brackets increases 
for smaller masses $M$. The excess spectrum in the mass region from 0.4 GeV 
to 0.75 GeV has been fitted by $w=0.19$. The slight downward shift 
about 40 MeV of the peak of the distribution compared with $m_{\rho }$ may
be estimated from Eq. (\ref{ll}) at these values of the parameters. This 
agrees with the above statement about small shift compared to $m_{\rho }$.
The total mass spectrum (the dashed line) and its widened Breit-Wigner 
component (the solid line) according to Eq. (\ref{ll}) with the chosen 
parameters are shown in Fig. 4. The overall description of experimental 
points seems quite satisfactory. The contribution of Cherenkov gluons (the 
excess of the dashed line over the solid one) constitutes the noticeable part 
at low masses. The formula (\ref{ll}) must be valid in the vicinity of the 
resonance peak. Thus we use it for masses larger than 0.4 GeV only.

From general principles one would expect 
slightly lower $p_T$ for low-mass dilepton pairs from coherent Cherenkov 
processes than for incoherent scattering. Qualitatively, this conclusion 
is supported by experiment \cite{da}. 

Whether the in-medium Cherenkov gluonic effect is as strong as shown in 
Fig. 4 can be verified by measuring the angular distribution of the lepton
pairs with different masses. The trigger-jet experiments similar to that at 
RHIC are necessary to check this prediction. One should measure the
angles between the companion jet axis and the total momentum of the lepton
pair. The Cherenkov pairs with masses between 0.4 GeV and 0.7 GeV
should tend to fill in the rings around the jet axis. The angular radius 
$\theta $ of the ring is determined by the usual condition. %(\ref{thet})

Another way to demonstrate it is to measure the average mass of lepton pairs
as a function of their polar emission angle (pseudorapidity) with the
companion jet direction chosen as $z$-axis. Some excess of low-mass pairs 
may be observed within the rings. 
Baryon-antibaryon effects can not possess signatures similar to these ones.

Some indications 
on the substructure with maxima at definite angles have been found at the 
same energies by CERES collaboration \cite{26}. It is not clear yet 
if it can be ascribed to Cherenkov gluons. To recover a definite maximum,
it would be better to detect a single parton jet, i.e. to have a trigger. 

The prediction of asymmetrical in-medium widening of {\bf any} resonance at its
low-mass side due to Cherenkov gluons is universal. This universality is 
definitely supported by experiment. Very clear signals of the excess on 
the low-mass sides of $\rho $, $\omega$ and $\phi$ mesons have been seen in 
\cite{5, Muto}. This effect for $\omega $-meson is also studied in
\cite{7}. Slight asymmetry of $\phi $-meson near 0.9 - 1 GeV is noticeable in
the Fig. 4 shown above but the error bars are large there. We did not try to 
fit it just to deal with as small number of parameters as possible. There are 
some indications at RHIC (see Fig. 6 in \cite{6}) on this effect for 
$J/\psi $-meson. However, the accuracy of RHIC data both for $\rho $ and
$J/\psi $ is not enough to get quantitative conclusions. 

To conclude, the new mechanism is proposed for explanation of the low-mass 
excess of dilepton pairs observed in experiment. It is the Cherenkov gluon 
radiation which adds to the ordinary processes at the left wing of any 
resonance. Only one adjustable parameter $w$ is used to describe its 
contribution to the dilepton spectra. The universal nature of their
asymmetry for {\bf all} resonances is predicted.

\section{Higher order effects leading to the "color rainbow"}

It is interesting to note that the general first order solutions of
Eqs. (\ref{f.6}), (\ref{f.7}), explicitly shown by Eqs. (\ref{f.12}), 
(\ref{f.13}) for Cherenkov gluons, stay valid at all orders if the nuclear 
permittivity does not depend on color. The higher order terms contribute
only if the permittivity differs for different colors. The general procedure 
of their calculation is the same as for in-vacuum QCD \cite{kovc, kov, ma}. 
After getting explicit lowest order solution one exploits
it together with the non-Abelian current conservation condition to find the
current component proportional to $g^3$. Then with the help of Eqs.
(\ref{f.6}), (\ref{f.7}) one finds the potentials up to the order $g^3$.
They can be represented as integrals convoluting the current with the
corresponding in-medium Green function. The even higher order corrections may 
be obtained in the same way. We outline briefly the path to the next order 
effects and qualitative results.

The third order terms of the potentials can be found \cite{adl} from the 
following expressions
\begin{eqnarray}
\epsilon (\bigtriangleup \Phi _a^{(3)}-\epsilon \frac 
{\partial ^2 \Phi _a^{(3)}}{\partial t^2})
=-gf_{abc}[\epsilon_f^2(\epsilon_fv^2-1)+(2\epsilon+\epsilon_f^2v^2)\Delta_{bc}]
 \frac {\partial \Phi _b^{(1)}}{\partial t}\Phi _c^{(1)},
\end{eqnarray}

$\Delta_{bc}=(\epsilon_b- \epsilon_c)/2 ;\;\;\;\;\;\;  \epsilon_f=(\epsilon_b+ \epsilon_c)/2$,

\begin{eqnarray}
\bigtriangleup A_{z,a}^{(3)}-\epsilon \frac{\partial ^2A_{z,a}^{(3)}}
{\partial t^2}=
-gf_{abc}[\epsilon_f(\epsilon_fv-1/v)+\Delta_{bc}(2v(\epsilon+\epsilon_f)-1/v)]
{\frac {\partial \Phi _b^{(1)}}{\partial t}}\Phi _c^{(1)}.
\end{eqnarray}

The density of the energy loss is proportional according to Eq. (\ref{eloss}) to
\begin{equation}
E_{z,a}^{(3)}=i\int \frac {d^4k}{(2\pi )^4}\left([\omega A_{z,a}^{(3)}({\bf k},
\omega )-k\zeta \Phi_a^{(3)}({\bf k},\omega )]-igf_{abc}\int \frac {d^4k'}
{(2\pi )^4}A_{z,b}^{(1)}(k')\Phi_c^{(1)}(k-k')\right).
\end{equation}

The last term corresponds to the truly non-abelian correction.
Only the linear terms in $\Delta _{bc}$ must be left everywhere because of the 
antisymmetry of $f_{abc}$. It demonstrates that the classical first order
solution stays valid for the color independent permittivity. $\Delta _{bc}$
can differ from 0 either due to dispersion effects or because of explicit
dependence of permittivity on color.

The calculation of the spectra proceeds \cite{adl} in the same way as it was 
done above for first order terms. 
For the sake of brevity, we reproduce here only the first term in the 
limit $\epsilon =\epsilon _f, \;\;\;\;\; v=1$:

\begin{equation}
\frac {dN^{(4)}_{a(A\Phi)lim}}{dldx d\omega }=\frac {3\alpha ^2 _S}{4{\sqrt x}}
f_{abc}Q_bQ_c{\rm Im}\left[\frac {(\epsilon -1)^{1/2}\Delta _{bc}}
{\epsilon ^2(\epsilon x-1)^{3/2}}\right ].
\end{equation}

The terms $\propto \frac{\nu {\sqrt x}}{[(x-x_0)^2+(\nu x_0)^2]^{5/4}}$ and
$\propto \frac{\nu } {{\sqrt x}[(x-x_0)^2+(\nu x_0)^2]^{3/4}}$ appear
which are different (albeit somewhat similar) to the lowest order terms.
Now, the shape is different from simple a'la Breit-Wigner form. The typical 
proportionality to $\Delta n$ is seen.

To conclude, one can say that in the case of $\Delta _{bc}\neq 0$ the 
non-abelian color quantum rainbow appears due to the higher order terms.

\section{The nuclear permittivity and the rest system of the nuclear matter}

In electrodynamics the permittivity of real substances depends on
$\omega $. Moreover it has the imaginary part determining the absorption.
E.g., ${\rm Re }\, \epsilon $ for water (see \cite{ja}) is approximately constant
in the visible light region ($\sqrt {\epsilon }\approx 1.34$), increases at
low $\omega $ and becomes smaller than 1 at high energies tending to 1
asymptotically. The absorption (${\rm Im }\, \epsilon $) is very small for 
visible light but dramatically increases in nearby regions both at low and 
high frequencies. Theoretically this behavior is ascribed to various collective
excitations in the water relevant to its response to radiation with different 
frequencies. Among them the resonance excitations are quite prominent (see, 
e.g., \cite{fe}). Moreover, the medium considered is at rest and the 
permittivity values are determined just in this frame. Even in electrodynamics, 
the quantitative theory of their energy dependence is still lacking, however.

Then, what can we say about the nuclear permittivity and the frame to define it
in?
 
The partons constituting high energy hadrons or nuclei interact during 
the collision for a very short time. Nevertheless, there are experimental 
indications that an intermediate state of matter (CGC, QGP, nuclear fluid ...)
is formed and evolves. Those are J/$\psi $-suppression, jet quenching, 
collective flow ($v_2$),
Cherenkov rings of hadrons etc. They show that there is collective response of 
the nuclear matter to color currents moving in it. Unfortunately, our
knowledge of its internal excitation modes is very scarce, much smaller than 
in electrodynamics. 

The permittivity is the internal property of a medium which demonstrates
the medium response to the {\it induced} current. Its quantitative 
description poses problems even in QED. It becomes more difficult task in
QCD where confinement is not understood. 
The attempts to calculate the nuclear permittivity from first principles are
not very convincing. It can be obtained from the polarization operator.
The corresponding dispersion branches have been computed in the lowest order
perturbation theory \cite{kk, kl, we}. Then the properties of collective 
excitations have been studied in the framework of the thermal field
theories (for review see, e.g., \cite{bi}). Their results with additional 
phenomenological ad hoc assumption about the role of resonances were used
in a simplified model of scalar fields \cite{ko} to show that the nuclear 
permittivity can be larger than 1 that admits Cherenkov gluons.

Our main goal is to study the medium response to the {\it external} color 
current. We prefer to use the general formulae of the scattering theory 
\cite{go} to estimate the nuclear permittivity. We have to rely on analogies 
and our knowledge of properties of hadrons. The only experimental
facts we get about this medium are brought by particles registered at the final 
stage. They have some features in common which (one can hope!) are also
relevant for gluons as the carriers of the strong forces. Those are the resonant
behavior of amplitudes at rather low energies and positive real part of the
forward scattering amplitudes at very high energies for hadron-hadron and
photon-hadron processes as measured from the interference of the Coulomb and
hadronic parts of the amplitudes. ${\rm Re} F(0^o,E)$ is always positive 
(i.e., $n>1$) within the low-mass wings of the Breit-Wigner resonances.  
This shows that the necessary condition for Cherenkov effects $n>1$ is
satisfied at least within these two energy intervals. This fact was used
to describe experimental observations at SPS, RHIC and cosmic ray energies. 
The asymmetry of
the $\rho $-meson shape at SPS \cite{da} and azimuthal correlations of 
in-medium jets at RHIC \cite{ul, ajit} were explained above by emission of 
comparatively low-energy Cherenkov gluons \cite{dnec, drem1}.
The parton density and intensity of the radiation were estimated. In its turn,
cosmic ray data \cite{apan} at energies corresponding to LHC ask for very high 
energy gluons to be emitted by the ultrarelativistic partons moving along the
collision axis \cite{d1}. Let us note the important difference from
electrodynamics where $n<1$ at high frequencies. For QGP the high-energy
condition $n>1$ is a consequence of its instability.

The in-medium equations are not Lorentz-invariant. There is no problem in
macroscopic electrodynamics because the rest system of the macroscopic matter
is well defined and its permittivity is considered there. For collisions of
two nuclei (or hadrons) it asks for special discussion.

Let us consider a particular parton which radiates in the nuclear matter.
It would "feel" the surrounding medium at rest if momenta of all other partons 
(or constituents of the matter), with which this parton can interact, 
sum to zero. In RHIC 
experiments the triggers which registered the jets (created by partons) 
were positioned at 90$^o$ to the collision axis. Such partons should be
produced by two initial forward-backward moving partons scattered at 90$^o$.
The total momentum of other partons (medium spectators) is balanced because 
for such geometry the partons from both nuclei play a role of spectators 
forming the medium. Thus the center
of mass system is the proper one to consider the nuclear matter at rest in
this experiment. The permittivity must be defined there. The Cherenkov rings 
consisting of hadrons have been registered around the away-side jet which
traversed the nuclear medium. This geometry requires however high statistics
because the rare process of scattering of initial partons at 90$^o$ has been 
chosen. 

The forward 
(backward) moving partons are much more numerous and have higher energies. 
However, one can not treat the radiation of such a primary parton in c.m.s.
in the similar way because the momentum of spectators is different from zero
i.e. the matter is not at rest. Now the spectators (the medium) are formed from 
the partons of another (target) nucleus only. Then the rest system of the medium
coincides with the rest system of that nucleus and the permittivity should 
refer to this system. The Cherenkov radiation of such highly energetic 
partons must be considered there. That is what was done for interpretation
of the cosmic ray event in \cite{d1}. This discussion clearly shows that 
one must carefully define the rest system for other geometries of the 
experiment with triggers positioned at different angles.

Thus our conclusion is that the definition of $\epsilon $ depends on the 
experiment geometry. Its corollary is that partons moving in different
directions with different energies can "feel" different states of matter in
the ${\bf same}$ collision of two nuclei because of the dispersive dependence
of the permittivity. The transversely scattered partons with comparatively 
low energies can analyze the matter with rather large permittivity corresponding
to the resonance region while the forward moving partons with high energies 
would "observe" low permittivity in the same collision. This peculiar feature
can help scan the $(\ln x, Q^2)$-plane as it is discussed in Section 5 and 
\cite {IJMP}. It explains also the different values of $\epsilon $ needed 
for description of RHIC and cosmic ray data.

 These conclusions can be checked at LHC because 
both RHIC and cosmic ray geometry will become available there. The energy
of the forward moving partons would exceed the thresholds above which 
$n>1$. Then both types of experiments can be done, i.e. the 90$^o$-trigger
and non-trigger forward-backward partons experiments. The predicted results for 
90$^o$-trigger geometry are similar to those at RHIC. The non-trigger
Cherenkov gluons should be emitted within the rings at polar angles of tens
degrees in c.m.s. at LHC by the forward moving partons (and symmetrically by 
the backward ones). This idea is supported by events observed in cosmic
rays \cite{apan, d1, drem1}. The experiments with triggers positioned at 
various angles to the collision axis should be done at the LHC.

\section{Conclusions}

The equations of in-medium gluodynamics (\ref{f.6}), (\ref{f.7}) are proposed. 
They remind the
in-medium Maxwell equations with non-Abelian terms added. Their lowest
order classical solutions are similar (up to the trivial color factors) 
to those of electrodynamics (\ref{f.12}), especially, for Cherenkov gluons. 

Some effects due to Cherenkov gluons at SPS, RHIC, cosmic rays and LHC energies 
have been discussed. The comparison with experimental data of RHIC allows to 
determine the real and imaginary parts of the nuclear permittivity.
It is related to the forward scattering hadronic amplitudes. 
This helps interpret the experimental results 
about the asymmetry of shapes of resonances traversing the nuclear medium.
Some indications on the energy dependence (dispersion) of the nuclear 
permittivity is obtained from comparison of RHIC and cosmic rays data.
This asks for the distinction between 
the different coordinate systems in which the Cherenkov radiation
(and nuclear permittivity) should be treated for partons moving in 
different directions with different energies.

Some estimates of properties of the nuclear matter formed in ultrarelativistic
heavy-ion collisions have been done. This consideration predicts new features 
at the LHC \cite{IJMP}. 

\bigskip

{\bf \large Acknowledgements}

\medskip

This work was supported
in parts by the RFBR grants 08-02-91000-CERN and 09-02-00741.

\newpage

{\bf Figure captions.}                                  \\
 
 Fig. 1. Away-side azimuthal correlations for STAR in the rapidity and 
 transverse momentum intervals indicated in the text and their description by
 the hypothesis about Cherenkov gluons.  \\

 Fig. 2. Away-side azimuthal correlations for PHENIX in the rapidity and 
 transverse momentum intervals indicated in the text and their description by
 the hypothesis about Cherenkov gluons.  \\

 Fig. 3. The distribution of the number of produced particles at different 
 distances from the event axis $r$ in the stratospheric event at 10$^{16}$ eV 
 \cite{apan} has two pronounced peaks. Correspondigly, the pseudorapidity 
 distribution possesses two such peaks.\\

 Fig. 4. Excess dilepton mass spectrum in semi-central In(158 AGeV)-In of NA60 (dots) 
compared to the in-medium $\rho $-meson peak with additional Cherenkov effect 
(dashed line).

\end{document}